\documentclass[a4paper,twoside]{article}
\newcommand{\affil}[1]{$^{\rm #1}$}
\usepackage[authoryear]{natbib}
\bibpunct{(}{)}{;}{a}{}{,}
\usepackage{graphicx}
\date{} 
\title{\large\bf\flushleft Time scales of the s process - from minutes to ages}
\author{\parbox{\textwidth}{\flushleft
\vspace{-0.5cm}
{\it F. K{\"a}ppeler\affil{A,F}, S. Bisterzo\affil{B}, R. Gallino\affil{B}, M. Heil\affil{C}, 
     M. Pignatari\affil{D}, R. Reifarth\affil{C}, O. Straniero\affil{E}, S. Walter\affil{A}, 
	 N. Winckler\affil{C}, K. Wisshak\affil{A}}\\
\vspace{0.4cm}
{\small \affil{A}\,Forschungszentrum Karlsruhe, Institut f{\"u}r Kernphysik, 
    P.O. Box 3640, D-76021 Karlsruhe, Germany}\\
{\small \affil{B}\,Dipartimento di Fisica Generale, Universit{\`a} di Torino, 
    Via P. Giuria 1, I-10125 Torino, Italy}\\
{\small \affil{C}\,GSI Darmstadt, Planckstr. 1, D-64291 Darmstadt, Germany}\\
{\small \affil{D}\,Keele University, Keele, Staffordshire ST5 5BG, UK and 
Joint Institute for Nuclear Astrophysics, University of Notre Dame, Notre Dame, IN 46556, USA}\\
{\small \affil{E}\,Osservatorio Astronomico di Collurania, I-64100 Teramo, Italy}\\
{\small \affil{F}\,Email: franz.kaeppeler@ik.fzk.de}}}
\begin{document}
\twocolumn[
\begin{changemargin}{.8cm}{.5cm}
\begin{minipage}{.9\textwidth}
\vspace{-1cm}
\maketitle
%
%
\small{\bf Abstract:}
A discussion of the time scales in the $s$ process appears 
to be an approriate aspect to discuss at the occasion of
Roberto's 70$^{th}$ anniversary, the more as this subject has
been repeatedly addressed during the 20 years of collaboration 
between Torino and Karlsruhe. The two chronometers presented in 
this text were selected to illustrate the intense mutual stimulation 
of both groups. Based on a reliable set of accurate stellar 
($n, \gamma$) cross sections determined mostly at FZK, the 
Torino group succeeded to develop a comprehensive picture of 
the various $s$-process scenarios, which are most valuable for
understanding the composition of the solar system as well
as for the interpretation of an increasing number of astronomical 
observations. 

\medskip
\medskip
\end{minipage}
\end{changemargin}
]
\small

%
\section{Introduction}

The question of time scales is directly related to a number of 
key issues in $s$-process nucleosynthesis and from the very 
beginning the corresponding chronometers were considered to 
represent an important source of information. The long-lived 
radioactivities, which are produced in the $s$ process, were 
studied first because of their cosmological importance and 
since they are less sensitive to details of stellar scenarios. 
But in principle any unstable isotope along the $s$-process
reaction path can be understood as a potential chronometer, 
provided that it defines a time scale corresponding to a 
significant quantity. 

The shortest possible time scale is related to {\it convective
mixing in He shell flashes}, which take place in thermally 
pulsing low mass asymptotic giant branch (AGB) stars, where 
turnover times of the less than an hour are attained (Sec. 
\ref{sec2}). 
  
{\it Neutron capture} occurs on time scales of days to years,
depending on the $s$-process scenario. The life time of a given 
isotope $A$ is determined by the stellar neutron flux, $n_n 
\times {\rm v}_T$, and by the stellar ($n, \gamma$) cross section,  
$$\tau_{n(A)} = \frac{1}{\lambda_{n(A)}} 
              = \frac{1}{n_n \times {\rm v}_T \times \sigma_{(A)}},$$
\noindent 
where $n_n$ denotes the neutron density and ${\rm v}_T$ the mean 
thermal velocity. If isotope $A$ is unstable against $\beta$-decay 
with a life time comparable to $\lambda_n$, the reaction path is
split into a branching with a characteristic abundance pattern that 
reflects this time scale and provides a measure for the neutron
density at the stellar site \citep{BSA01,WVA01}. This situation
can be complicated by the fact that in most cases only theoretical 
evaluations are available for the neutron capture cross sections
of the unstable isotopes. Furthermore, the beta decay the $\beta$-decay 
rate of the branch point isotope $A$ may depend on temperature and/or 
electron density of the stellar plasma as in case of $^{176}$Lu
(Sec. \ref{sec3}).

If the neutron capture time is comparable to the {\it duration of 
the neutron bursts}, the abundance pattern of such a branching can 
be used to test the time scale of the He shell flashes in AGB stars. 
Suited branchings of this type could be those at $^{85}$Kr 
and $^{95}$Zr with neutron capture times of a few years. In these 
cases, the neutron flux dies out before reaction equilibrium is 
achieved. However, both branchings are difficult to analyze because 
of significant contributions from the weak $s$-process component 
from massive stars or from the $r$ process, and a constraining 
analysis is difficult to achieve.

The {\it transport to the stellar surface} in the third dredge-up
phase can be investigated by means of the observed Tc abundances 
(Mathews et al. 1986; Smith and Lambert 1988;
Busso et al. 1995). Analyses of these observations have to 
consider that the terrestrial decay rate of $^{99}$Tc ($t_{1/2} = 
2.1 \times 10^5$ years) is reduced to a few years at $s$-process 
temperatures \citep{Sch83,TaY87}, which implies that it is quickly 
cooled to temperatures below 10$^8$ K after its production. A 
complementary, temperature-independent chronometer for the third 
dredge-up is $^{93}$Zr, which can be followed by the appearance 
of its daughter $^{93}$Nb \citep{MTW86}.

The time scale for the {\it formation of the solar system} can, in 
principle, be inferred from the abundance patterns, which are affected
by the decay of nuclei with half-lives between 10$^5$ and 10$^7$
years \citep{Pag90}. Quantitative studies based on isotopic
anomalies found in presolar grains have confirmed that such
effects exist for $^{26}$Al, $^{41}$Ca, $^{60}$Fe, $^{93}$Zr, 
and $^{107}$Pd. A comprehensive overview of this discussion was
presented by Busso, Gallino, \& Wasserburg (1999) (see 
also the contribution by R. Reifarth to this volume). Also $^{205}$Pb was discussed as a
potentially promising chronometer (Yokoi, Takahashi, \& Arnould 1985), whereas $^{53}$Mn, 
$^{129}$I, and $^{182}$Hf were found to result from the continuous
pollution of the interstellar medium by explosive nucleosynthesis 
in supernovae (Busso et al. 1999).

A number of attempts have been made to constrain the {\it cosmic
time scale} by means of $s$-process abundance information. In the
course of these studies it turned out that the half-life of the
most promising case, $^{176}$Lu (Audouze, Fowler, \& Schramm 1972; Arnould 1973), was strongly
temperature-dependent, making it an $s$-process thermometer rather 
than a cosmic clock as discussed in Sec. \ref{sec3}. The other 
long-lived species, $^{40}$K \citep{BeP87} and $^{87}$Rb (Beer and Walter 1984)
are produced by at least two different processes and are difficult 
to interpret quantitatively. Therefore, recent analyses of nuclear 
chronometers for constraining the cosmic time scale concentrate 
on the $r$-process clocks related to the decay of the long-lived
actinides \citep{CPK99} and of $^{187}$Re (Yokoi et al. 1983; 
Arnould, Takahashi, \& Yokoi 1984; Mosconi et al. 2007; see 
also the contribution by A. Mengoni to this volume). 

In the following sections we will focus on the shortest $s$-process 
time scale related to the fast convective mixing during the He shell
flashes in AGB stars.

\section{The branching at $^{128}$I \label{sec2}}

\begin{figure}[ht]
\begin{center}
\includegraphics[scale=0.78, angle=0]{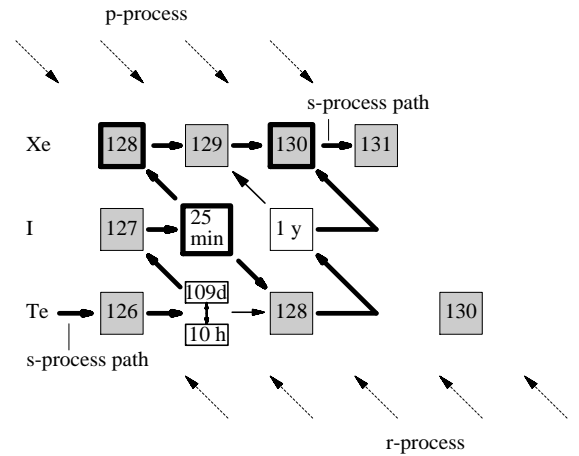}
\caption{The $s$-process reaction path between Te and Xe. The isotopes 
  $^{128}$Xe and $^{130}$Xe are shielded against $r$-process contributions by 
  their stable Te isobars. In contrast to $^{130}$Xe, $^{128}$Xe is partly 
  bypassed due to the branching at $^{128}$I. The branching at $^{127}$Te is 
  negligible unless the temperature is low enough that ground state and isomer 
  are not fully thermalized. The branching at $^{128}$I is unique since it 
  results from the competition between ${\beta^-}$ and electron capture decays 
  and is, therefore, independent of the neutron flux. 
\label{fig1}}
\end{center}
\end{figure}

Xenon is an element of considerable astrophysical interest.  
The origin of the lightest isotopes, $^{124}$Xe and $^{126}$Xe, can
be exclusively ascribed to the so-called $p$ process in supernovae
(Arnould and Goriely 2003). Their relative isotopic abundances are 
important for testing $p$-process models describing the proton-rich 
side of the valley of stability. Concerning the $s$ process, xenon 
belongs to the six elements with a pair of $s$-only isotopes. In 
this case, the relevant nuclei are $^{128}$Xe and $^{130}$Xe, both 
shielded against the decay chains from the $r$-process region by 
their stable Te isobars. The abundances of these isotopes define 
the strength of the branching in the $s$-process reaction chain 
illustrated in Fig.~\ref{fig1}. Since the $p$-process components 
of these $r$-shielded nuclei do not exceed a few percent, they are 
commonly considered to be of pure $s$ origin. On the neutron-rich 
side, $^{134}$Xe and $^{136}$Xe can be ascribed to the $r$ process
since the $\beta^-$ half life of $^{133}$Xe is short enough 
to prevent any significant $s$-process contributions. Hence, the Xe 
isotope chain carries signatures of all nucleosynthesis scenarios 
that contribute to the mass region of the heavy isotopes with 
$A\geq$ 90, and offers the possibility to constrain the underlying 
models. A detailed description of the isotopic abundance pattern 
of xenon involves necessarily quantitative models for all these 
processes. 

The combined strength of the branchings in the $s$-process chain at 
$^{127}$Te and $^{128}$I (Fig. \ref{fig1}) is defined by the relative 
abundances of the $s$-only isotopes $^{128}$Xe and $^{130}$Xe. Both 
branchings are expected to be comparably weak since only a small part 
of the total reaction flow is bypassing $^{128}$Xe. Therefore, the 
$\langle\sigma\rangle N_s$ value for $^{128}$Xe, which is characteristic 
of the reaction flow, is slightly smaller than the one for $^{130}$Xe. 
Since the solar isotopic ratio of the $s$-only isotopes is well defined 
(Pepin, Becker, \& Rider 1995), the  $\langle\sigma\rangle N_s $ difference 
can be obtained by an accurate measurement of the cross section ratio 
(Reifarth et al. 2002). 

While the first branching at $^{127}$Te is marginal because the 
population of ground state and isomer is quickly thermalized in the hot 
stellar photon bath (Takahashi and Yokoi 1987), leading to a strong 
dominance of the $\beta$-decay channel, the second branching at 
$^{128}$I is utterly interesting. In contrast to all other relevant 
cases, this branching is only defined by the competition between 
$\beta^-$ and electron capture decay (Figure \ref{fig1}). Both 
decay modes are sufficiently short-lived that the neutron capture 
channel and hence the influence of the stellar neutron flux is 
completely negligible. 

Since the electron capture rate is sensitive to temperature and 
electron density of the stellar plasma \cite{TaY87}, this branching 
provides a unique possibility to constrain these parameters without 
interference from the neutron flux. Under stellar conditions the 
corresponding branching ratio at $^{128}$I is
$$    
f_- = \lambda_{\beta^{-}} / (\lambda_{\beta^{-}}+\lambda_{\beta^{EC}}) = 
1-\lambda_{\beta^{EC}} / (\lambda_{\beta^{-}}+\lambda_{\beta^{EC}}).$$
While the $\beta^-$-rate varies only weakly, the electron capture 
rate depends strongly on temperature due to the increasing degree 
of ionization. Furthermore, at high temperatures, when the ions are 
fully stripped, the EC rate becomes sensitive to the density in 
the stellar plasma via electron capture from the continuum (Table 
\ref{tab2}). The relatively small change of the branching ratio
did not permit quantitative analyses until the stellar ($n, \gamma$) 
rates of the involved isotopes were measured to an accuracy of 
1.5\% \cite{RHK02}.

\begin{table}[h]
\caption{Beta-decay branching ratio $f_-$ at $^{128}$I as a function of
         electron density $n_e$ and temperature \cite{TaY87}. \label{tab2}}
\begin{tabular}{ccccc}
\hline		  		  	  			\\
$n_e$ ($10^{26}$ cm$^{-3}$) &
\multicolumn{4}{c}{Temperature (10$^8$ K)}\\
   & 0     & 1     & 2     & 3     \\
\hline	   						   \\
0  & 0.940 & 0.963 & 0.996 & 0.999 \\
3  & 0.940 & 0.952 & 0.991 & 0.997 \\
10 & 0.940 & 0.944 & 0.976 & 0.992 \\
30 & 0.940 & 0.938 & 0.956 & 0.980 \\
\hline	   	 	   	 	   	 	   \\
\end{tabular}
\end{table}

\subsection{Thermally pulsing AGB stars}

The $s$-process abundances in the mass range $A \ge$ 90 are 
produced during helium shell burning in thermally pulsing low 
mass AGB stars \cite{SGB95} by the subsequent operation of two 
neutron sources. The $^{13}$C($\alpha, n$)$^{16}$O reaction, which
occurs under radiative conditions at low temperatures ($kT \approx$8 
keV) and neutron densities ($n_n \leq 10^7$ cm$^{-3}$) between 
convective He-shell burning episodes, provides most of the 
neutron exposure. The resulting abundances are modified
by a second burst of neutrons from the $^{22}$Ne($\alpha, 
n$)$^{25}$Mg reaction, which is marginally activated during the 
highly convective He shell flashes, when peak neutron 
densities of n$_n \ge 10^{10}$ cm$^{-3}$ are reached at 
$kT \approx$23 keV. Although this second neutron burst 
accounts only for a few percent of the total neutron exposure, 
it is essential for adjusting the final abundance patterns of 
the $s$-process branchings. It is important to note 
that the ($n, \gamma$) cross sections in Te-I-Xe region are large 
enough that typical neutron capture times are significantly shorter 
than the duration of the two neutron exposures. Therefore, the 
abundances can follow the time variations of the neutron density.

\subsection{Convection in He shell flashes}

\begin{figure}
\begin{center}
\includegraphics[scale=0.45, angle=0]{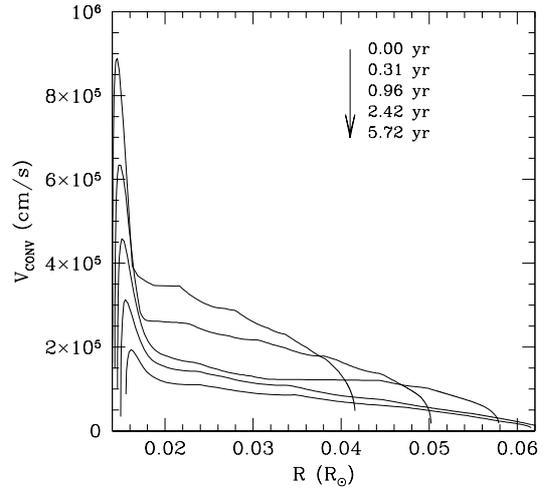}
\caption{The calculated convective velocities as a function of the internal 
radius for 5 models ($t$=0, 0.31, 0.96, 2.42 and 5.72 yr after the maximum 
of the TP) in a 3 $M_{\odot}$ star of solar composition. The convective turnover 
time is a few hours only. The scale on the abscissa starts at the bottom of the 
convective shell.
\label{fig2}}
\end{center}
\end{figure}

The effect of convection for the branchings at $A=127/128$ was
extensively studied by means of the stellar evolution code FRANEC 
(Straniero et al. 1997; Chieffi and Straniero 1989) using a time-dependent 
mixing algorithm to treat the short time scales during thermal pulses 
along the AGB. Convective velocities were evaluated by means of the mixing 
length theory where the mixing length parameter was calibrated by a fit of 
the solar radius. An example for these results is given in Fig. \ref{fig2}, 
which represents a typical thermal pulse for a 3 $M_{\odot}$ AGB star 
of solar composition. The calculated convective velocities 
are plotted for five models ($t$=0, 0.31, 0.96, 2.42 and 5.72 yr 
after the maximum of the TP) as a function of the internal radius
to show how the convective shell expands after the pulse maximum.  

The rather short convective turnover times of less than one hour that can be 
derived from Fig.\,\ref{fig2} are in any case, shorter than the time during 
which the temperature at the bottom of the convective pulse remains higher 
than $2.5 \times 10^8$ K. Hence, the crucial transport time from the hot 
synthesis zone to cooler layers is only of the order of minutes. This can 
have an impact on those potential $s$-process branchings, which are 
characterized by branch-point isotopes with strong thermal enhancements of 
their decay rates. Even if such half-lives are reduced to a few minutes at 
the bottom of the He shell flash, the signature of the branching can survive
by the rapid mixing of processed material into the cooler outer layers of
the convective zone. 

\subsection{Branching analysis at {\emph A}=127/128}

The ($n, \gamma$) cross sections required for describing the time evolution 
of the $^{128}$Xe and $^{130}$Xe abundances during the He shell flashes have 
been accurately measured \cite{RHK02,ReK02}. This information is crucial
in view of the fact that only a comparably small fraction of the 
reaction flow is bypassing $^{128}$Xe (Table \ref{tab2}). With the Xe 
cross sections compiled by Bao et al. (2000) for example, the $^{128}$Xe/$^{130}$Xe 
ratio could only be calculated with an uncertainty of about 20\%, whereas
this uncertainty was reduced to $\pm1.5$\% with the improved cross section 
data. Another source of uncertainty in the nuclear input is due to the
theoretical $\beta$-decay rates listed in Table \ref{tab2}. Variation of these 
rates by a factor of two affects the branching ratio by less than 2\%,
because the decay of $^{128}$I is dominated by the $\beta^-$-mode.  

In fact, the branching analysis indicates a more complex
situation than one might expect from the trends of the branching factor
in Table \ref{tab2}. During the low temperature phase between He shell 
flashes, the neutron density produced via the $^{13}$C($\alpha$, n)$^{16}$O 
reaction is less than 10$^7$ cm$^{-3}$. The branching at $^{127}$Te is completely 
closed, resulting in a $^{128}$Xe/$^{130}$Xe abundance ratio of 0.93 relative
to the solar values at the end of the low temperature phase due to the effect 
of the $^{128}$I branching. After the onset of convection at the beginning 
of the He shell flash the production factors of $^{128}$Xe and $^{130}$Xe 
differ by 8\%, corresponding to the solar ratio, but are then modified 
during the flash. Since the ($n, \gamma$) cross sections of both isotopes 
are large enough for achieving local $\langle\sigma\rangle N_s $ 
equilibrium, one would expect that the production factor of $^{128}$Xe quickly 
approaches that of $^{130}$Xe. In this case, the final abundance ratio 
would clearly exceed the solar value.

In contrast, one finds that the branching still exists even 
during the high temperature phase of the He shell flash. There are 
essentially two effects, which concur to explain this behavior:
\begin{itemize}
\item During the peak of temperature and neutron density, the electron 
densities at the bottom of the convective He shell flash, i.e. in the 
$s$-process zone, are between 15 $\times$ 10$^{26}$ cm$^{-3}$ and 20 
$\times$ 10$^{26}$ cm$^{-3}$. Therefore, the branching at $^{128}$I 
is never completely closed. Even at the peak temperatures of the 
He-shell flash, typically 3\% of the flow are bypassing $^{128}$Xe.
\item Around the maximum of the neutron density, the branching at 
$^{127}$Te is no longer negligible and leads to an additional
decrease of the $^{128}$Xe abundance. 
\end{itemize}

In view of these effects the $^{127}$Te branching 
can be fully understood only if the stellar reaction network is 
calculated with sufficiently small time steps so that   
the time scale of convective turnover is properly considered. If the neutron density 
is followed in time steps of 10$^5$ s up to freeze-out one finds an enhancement of
the $^{127}$Te branching of 30\% compared to calculations performed 
on a time grid of 10$^6$ s. However, reducing the time steps 
to $3\times10^4$ s had no further effect on the abundance ratios. 

If averaged over the AGB evolution, the stellar models yield a final 
abundance ratio of $^{128}$Xe/$^{130}$Xe = 0.466$\pm$0.015, 
consistent with the smaller value measured in the solar system = 
0.510$\pm$0.005 if an 8\% $p$ contribution to the solar $^{128}$Xe 
abundance is taken into account. 

In summary, the abundance ratio of $^{128}$Xe and $^{130}$Xe in the solar 
system could be eventually reproduced by combining the effects of 
mass density, temperature, neutron density, and convective turnover  
in a consistent way. This success represents an impressive confirmation 
of the stellar $s$-process model related to thermally pulsing low mass 
AGB stars.

\section{The $^{176}$Lu/$^{176}$Hf pair \label{sec3}}

The mass region of the rare earth elements (REE) represents an 
important test ground for $s$-process models because the relative 
REE abundances are known to better than $\pm$2\%, which implies 
that the branchings in this region are reliably defined by the 
respective $s$-only isotopes. Systematic analyses of these branchings
contribute essentially to the quantitative picture of the
main $s$-process component that has been achieved with stellar
models for thermally pulsing low mass AGB stars (Gallino et al. 
1998; Busso et al. 1999). Also the stellar ($n, \gamma$) cross 
sections (except for the lightest REE) are large enough that the 
assumption of reaction flow equilibrium is well justified with 
respect to the analysis of several prominent $s$-process branchings 
in this mass region. This means that the final abundances are 
essentially determined during the freeze-out phase at the decline 
of neutron density towards the end of the shell flash.

\begin{figure}[t]
\begin{center}
\includegraphics[width=0.45\textwidth]{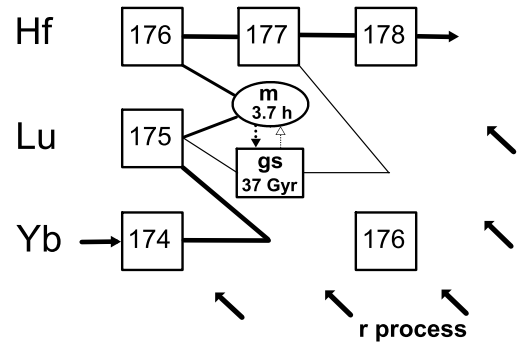}
\caption{The $s$-process reaction flow in the Lu region. The 
strength of the lines indicates that neutron captures on 
$^{175}$Lu leading to the isomeric state in $^{176}$Lu are 
more probable than those to the ground state.\label{fig3}}
\end{center}
\end{figure}

Among the REE branchings, $^{176}$Lu is especially attractive 
for the intricate way in which nuclear physics can affect the
actual $s$-process production yields. This is illustrated in
Fig.\,\ref{fig3} showing that the reaction path in the vicinity
of lutetium is determined not only by the stellar neutron capture
cross sections of $^{175,176}$Lu and $^{176}$Hf, but also by the
thermal coupling of isomer and ground state in $^{176}$Lu.

Due to its long half-life of 37.5 Gyr, $^{176}$Lu was initially
considered as a potential nuclear chronometer for the age of the 
$s$ elements \citep{AFS72}. This possibility is based on the fact 
that $^{176}$Lu as well as its daughter $^{176}$Hf are of pure 
$s$-process origin, since both are shielded against the
$r$-process beta decay chains by their stable isobar $^{176}$Yb. In
a straightforward approach, the reaction flow in the branching at $A
= 176$ and, therefore, the surviving $s$ abundance of $^{176}$Lu and
of $^{176}$Hf would be determined by the partial ($n, \gamma$) cross
sections of $^{175}$Lu feeding the ground and isomeric state of
$^{176}$Lu. Since transitions between the two states are highly
forbidden by selection rules, both states could be considered as 
separate species in the description of the $s$-process branching at 
$A$ = 176 (Fig.~\ref{fig4}). While the isomer rapidly decays quickly
($t_{1/2} = 3.7$ h) by $\beta$-transitions to $^{176}$Hf, the 
effective $s$-process yield of $^{176}$Lu appeared well defined 
by the partial cross section to ground state, thus providing an 
estimate the age of the $s$ elements by comparison with the actual 
solar system value.

However, Ward and Fowler (1980) noted that ground state and isomer 
of $^{176}$Lu are most likely connected by nuclear excitations in 
the hot stellar photon bath, since thermal photons at $s$-process
temperatures are energetic enough to populate higher lying states, 
which can decay to the long-lived ground state and to the short-lived 
123 keV isomer as well. In this way, the strict forbiddeness of direct 
transitions between both states is circumvented, dramatically reducing 
the effective half-life to a few hours. As a result, most of the 
reaction flow could have been directly diverted to $^{176}$Hf, 
resulting in a $^{176}$Lu abundance much smaller than observed in 
the solar system. That this temperature dependence had indeed affected 
the $^{176}$Lu/$^{176}$Hf abundance ratio was confirmed soon 
thereafter by Beer et al. (1981, 1984), thus changing its 
interpretation from a potential chronometer into a sensitive 
$s$-process thermometer.

The temperature dependence was eventually quantified on the basis
of a comprehensive investigation of the level structure of
$^{176}$Lu (Klay et al. 1991b; Klay et al. 1991a; Doll et al. 
1999). The lowest mediating state was identified at an excitation 
energy of 838.6 keV, which implies that thermally induced transitions 
become effective at temperatures above $T_8 = 1.5 -2$ (where $T_8$ 
is the temperature in units of 10$^8$ K). Accordingly, ground state 
and isomer can be treated as separate species only at lower 
temperatures. In thermally pulsing low mass AGB stars this is the 
case between convective He shell flashes when the neutron production 
is provided by the $^{13}$C($\alpha, n$)$^{16}$O reaction in the
so-called $^{13}$C pocket. Under these conditions the abundance 
of $^{176}$Lu is directly determined by the partial cross sections 
populating ground state and isomer. 

During the He shell flashes, the higher temperatures at the
bottom of the convective region lead to the activation of
the $^{22}$Ne($\alpha, n$)$^{25}$Mg reaction. It is in this 
regime that the initial population of ground state and isomer
starts to be changed by thermally induced transitions. 
This affects the $^{176}$Lu$^{\rm g}$ in the long-lived 
ground state that is actually produced by neutron capture 
in the burning zone as well as the Lu fraction circulating 
in the convective He shell flash. The latter part is exposed 
to the high bottom temperature only for rather short times and, 
therefore, less affected by the temperature dependence of the 
half life. Once produced, by far most of the long-lived 
$^{176}$Lu$^{\rm g}$ survives in the cooler layers outside of 
the actual burning zone. 

\subsection{Nuclear physics input}

The stellar neutron capture rates for describing the reaction 
flow in Fig.~\ref{fig3} were determined by accurate time-of-flight 
(TOF) measurements of the total ($n, \gamma$) cross sections for
$^{175}$Lu and $^{176}$Lu using a 4$\pi$ BaF$_2$ array (Wisshak et 
al. 2006). The Maxwellian averaged cross sections (MACS) deduced 
from these data are five times more accurate than the values listed 
in the compilation of Bao et al. (2000). However, neutron captures 
on $^{175}$Lu may feed either the ground state or the isomer in 
$^{176}$Lu. Therefore, the total ($n, \gamma$) cross section was 
complemented by a measurement of at least one of the two partial 
cross sections ($\sigma_{p}^{\rm g}$, $\sigma_{p}^{\rm m}$). Since 
the corresponding reaction channels could not be distinguished in 
the TOF measurement \citep{WVK06a}, the activation technique was
used to determine the partial cross section to the isomeric state 
in $^{176}$Lu at thermal energies of $kT = 5.1$ and 25 keV.

Activation measurements of the partial ($n, \gamma$) cross section 
to the isomeric state in $^{176}$Lu at or near a neutron energy 
of 25 keV were performed via the $^7$Li($p, n$)$^7$Be reaction 
(Beer and K{\"a}ppeler 1980; Allen, Lowenthal, \& de~Laeter 1981;
Zhao and K{\"a}ppeler 1991) and also using a filtered neutron beam 
from a nuclear reactor (Stecher-Rasmussen et al. 1988). In all
experiments the induced activity after irradiation was detected via
the 88 keV $\gamma$-transition in the decay of $^{176}$Lu$^{\rm m}$.
These data were recently complemented by an activation at $kT = 5$ keV,
adapted to the lower temperature in the $^{13}$C pocket \cite{HWD08}, 
where most of the neutron exposure is provided by the $^{13}$C($\alpha, 
n$)$^{16}$O source, which operates at $kT = 8$ keV.

\begin{figure}[t]
\begin{center}
\includegraphics[width=0.4\textwidth]{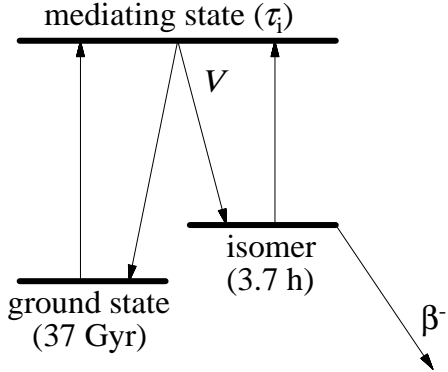}
\end{center}
\caption{Schematic level scheme of $^{176}$Lu illustrating the
thermal coupling between the long-lived ground state and the
short-lived isomer. While direct transitions between these states
are strictly forbidden by selection rules, thermal excitations in
the hot stellar photon bath populate a higher lying state with
intermediate quantum numbers that can decay both ways. At first this
link depopulates the isomer towards the ground state because it is
more easily reached from the isomer. However, at sufficiently high
temperatures, induced transitions from the ground state to the
mediating state can also feed the isomer, resulting in the
destruction of $^{176}$Lu via $\beta$ decay of the short-lived
isomer. The efficiency of this mechanism is determined by the
lifetime $\tau_i$ of the mediating state and by the branching ratio
$V$ for decays towards the isomer. \label{fig4}}
\end{figure}

The half-life of $^{176}$Lu was shown to decrease drastically
at the temperatures reached during He shell burning in
thermally pulsing low mass AGB stars (Klay et al. 1991a;
Doll et al. 1999). Contrary to the situation at low excitation
energy, where transitions between ground state and isomer are
strictly forbidden by selection rules, interactions with
the hot stellar photon bath lead to induced transitions to
higher lying nuclear states, which can decay into the ground
state and into the isomer as well (Fig. \ref{fig4}). 

Up to $T_8 = 2$ the mediating states can not be reached and
the reaction path of Fig. \ref{fig3} is completely defined
by the partial capture cross sections feeding ground state
and isomer in $^{176}$Lu. This situation prevails in the 
$^{13}$C pocket between He shell flashes. Between $T_8 =2.2$ 
and 3.0, ground state and isomer start to be increasingly 
coupled. At first, this coupling leads to an increasing 
population of the ground state, because of the smaller energy
difference between isomer and mediating state. It is due to 
this effect that more $^{176}$Lu is observed in nature than
would be created in a "cool" environment with $T_8 \leq
2$. In this regime, internal transitions, $\beta$-decays, and
neutron captures are equally important and have to be properly
considered during He shell flashes, where the final abundance
pattern of the $s$-process branchings are established by the 
marginal activation of the $^{22}$Ne($\alpha, n$)$^{25}$Mg 
neutron source reaction at thermal energies around $kT$ = 
23 keV. 

\subsection{Branching factor}

The branching factor $f_n$ describing the split of the reaction flow
at $^{176}$Lu (Fig. \ref{fig3}) can be expressed in terms of the
neutron capture rate $\lambda_n = \langle \sigma_{176} \rangle v_T
n_n$ (where $\sigma_{176}$ denotes the MACS of $^{176}$Lu), $v_T$
the mean thermal neutron velocity, and $n_n$ the neutron density)
and of the temperature and neutron density dependent $\beta$-decay
rate of $^{176}$Lu. Because of the thermal effects on the lifetime
of $^{176}$Lu sketched above, $f_n = \lambda_n / (\lambda_n +
\lambda_\beta$) becomes a complex function of temperature and
neutron density (Klay et al. 1991a; Doll et al. 1999).

The bottom temperature in the He flashes reaches a maximum at the
maximum expansion of the convection zone and then declines rapidly
as the convective thermal instability shrinks. According to current 
AGB models of low mass stars, the neutron burst released by the 
$^{22}$Ne neutron source lasts for about six years, as long as 
$T_8 \ge$ 2.5, while the convective instability lasts for a much 
longer period of about 150 years. The maximum temperature in the 
He flash increases slightly with pulse number, reaching $T_8 
\approx$ 3 in the more advanced pulses \citep{SDC03}.

Although the second neutron burst during the He shell flashes
contributes only a few percent to the total neutron exposure, the
final composition of the Lu/Hf branching is, in fact, established
during the freeze-out phase at the end of the thermal pulse. It is 
only during the high-temperature phase that the thermal coupling 
between ground state and isomer takes place. In this context, it 
is also important to remember that the ($n, \gamma$) cross sections 
in the Lu/Hf region are large enough that typical neutron capture 
times are significantly shorter than the duration of the neutron 
exposure during the He shell flash.

Previous $s$-process calculations for thermally pulsing low mass 
AGB stars (Gallino et al. 1988; Gallino et al. 1998), which were 
performed by post-processing using full stellar evolutionary
models obtained with the FRANEC code (Chieffi and Straniero 1989;
Straniero et al. 1995; Straniero et al. 2003; Straniero, Gallino, 
\& Cristallo 2006) were successful in describing the solar main 
$s$-process component fairly well (Gallino et al. 1998; Arlandini 
et al. 1999). For the treatment of the branching at $^{176}$Lu, 
however, this approach had to be refined by subdividing the 
convective region of the He shell flashes into 30 meshes in order to 
follow the neutron density profile in sufficient detail and to take 
the strong temperature dependence of the branching factor $f_n$ 
properly into account. The production and decay of $^{176}$Lu was 
calculated in each mesh to obtain an accurate description of the 
final $^{176}$Lu/$^{176}$Hf ratio in the He shell flash.

After each time step of less than one hour, which corresponds to 
the typical turnover time for convection in the He shell flash
\citep{RKV04}, the abundances from all zones were averaged in order
to account for the fast convective mixing. This treatment is well
justified because there is no efficient coupling between ground
state and isomer at temperatures below $T_8 \leq 2.5$, which holds
for all meshes except the first few from the bottom. Interestingly,
this is the temperature when neutron production via the
$^{22}$Ne($\alpha, n$)$^{25}$Mg reaction diminishes.

\begin{figure}[t]
\includegraphics[width=0.35\textwidth, angle=270]{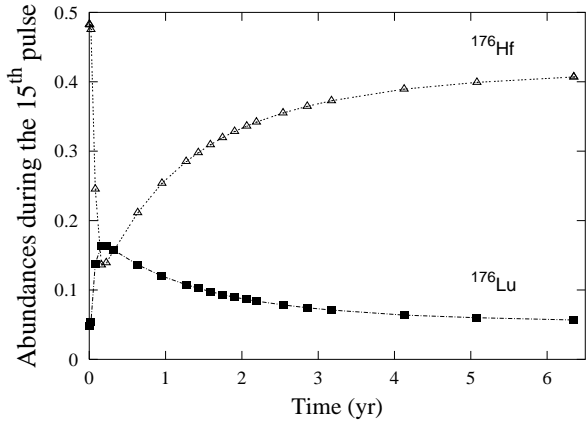}
\caption{
Evolution of the $^{176}$Lu (squares) and $^{176}$Hf (circles) abundances
during the 15$^{\rm th}$ He shell flash in an AGB star with 1.5 $M_{\odot}$. 
The time scale starts when the temperature at the bottom of the convective 
shell reaches $T_8 = 2.5$, i.e. at the onset of the $^{22}$Ne($\alpha$,
n)$^{25}$Mg reaction. The values are given as number fractions normalized
to that of $^{150}$Sm at the end of the He shell flash.
\label{fig5}}
\end{figure}

During the main neutron exposure provided by the $^{13}$C($\alpha,
n$)$^{16}$O reaction thermal effects are negligible for the
production of $^{176}$Lu because the temperatures in the $^{13}$C
pocket of $T_8 \approx 1$ are not sufficient to reach the mediating
state at 838.6 keV. Accordingly, the $^{176}$Lu/$^{176}$Hf ratio is
simply determined by the partial ($n, \gamma$) cross section to the
ground state, which would always produce too much $^{176}$Hf and too
little $^{176}$Lu$^{\rm g}$. This situation is illustrated in Fig.
\ref{fig5} showing the relative production factors of $^{176}$Lu and
$^{176}$Hf during the 15$^{\rm th}$ He shell flash of the AGB model
with initial mass 1.5 $M_{\odot}$ and half-solar metallicity, which
is representative of the overall relative abundance distribution
ejected during the AGB phase. The $s$-process abundances are given 
as number fractions $Y_i = X_i / A_i$ ($X$ being the respective mass 
fractions), normalized to that of $^{150}$Sm at the end of the He 
shell flash. This isotope was chosen as a reference because it is 
the best known case of an unbranched $s$-only isotope among the REE.

After the $s$-process nuclides synthesized in the $^{13}$C
pocket are engulfed by the convective He shell, one finds
the ratio shown at $t = 0$ in Fig. \ref{fig5}, which marks
the onset of $s$ process nucleosynthesis in the He shell flash.
In this phase temperatures are high enough to populate the 
the mediating state at 838 keV, leading to a strong increase 
in the production of the long-lived ground state of 
$^{176}$Lu$^{\rm g}$. As shown in Fig. \ref{fig5} $^{176}$Hf 
is almost completely destroyed at the beginning of the flash,
when neutron density and temperature are highest and $^{176}$Hf 
is efficiently bypassed by the reaction flow. Correspondingly,
the $^{176}$Lu abundance reaches a pronounced maximum. As
temperature and neutron density decline with time, the
branching towards $^{176}$Hf is more and more restored, but
the final abundance at the end of the He shell flash remains
significantly lower than at the beginning. Because the thermal 
coupling between ground state and isomer depends so critically 
on temperature, the full $s$-process network was followed in 
each of the 30 meshes of the convective zone. 

Plausible solutions of the Lu/Hf puzzle are characterized by 
equal overproduction factors for $^{176}$Lu and $^{176}$Hf, at 
least within the experimental uncertainties of the nuclear input
data. The decay of long-lived $^{176}$Lu ground state in the 
interstellar medium prior to the formation of the solar system
reduces the calculated overproduction factor of $^{176}$Lu
by about 10\%, leaving the more abundant $^{176}$Hf essentially
unchanged. In the light of these considerations, overproduction 
factors between 1.00 and 1.10 are acceptable for $^{176}$Lu and 
between 0.95 and 1.05 for $^{176}$Hf.

In the course of these investigations it turned out that the 
observed $^{176}$Lu/$^{176}$Hf ratio could only be reproduced if 
the thermal coupling of ground state and isomer in $^{176}$Lu
in combination with the neutron density was followed within the 
gradients of time, mass, and temperature by the refined zoning 
of the convective He shell flash. Only in this way overproduction 
factors could be obtained, which were compatible with the limits 
defined by the nuclear physics uncertainties and by the $^{176}$Lu 
decay. In contrast to a previous study \citep{AKW99b}, the 
meanwhile improved cross section information together with the 
multi-zone approach for the He shell flash appears to settle the 
$^{176}$Lu puzzle. While the $^{176}$Hf abundance was before 
overproduced by 10 $-$ 20\%, the present calculations yield final 
$^{176}$Lu$^{\rm g}$ and $^{176}$Hf abundances (relative to solar 
system values) of 1.04 and 0.96. These results do not yet consider 
the decay of the produced $^{176}$Lu$^{\rm g}$, a correction that 
brings the two numbers even into closer agreement.

Recently, additional information from Coulomb excitation 
measurements \citep{VLP00} and photoactivation studies \citep{Kne05} 
have triggered renewed interest in the Lu/Hf problem 
\citep{Moh08} and seem to provide further constraints
for the convection at the bottom of the He shell flash.

\section{Conclusions}

The elemental and isotopic abundances found in nature carry 
the signatures of their nucleosynthetic origin as well as of 
the related time scales. The branchings at $^{128}$I and 
$^{176}$Lu discussed in Secs. \ref{sec2} and \ref{sec3} are 
characterized by the rather short time scale of convection 
in He shell flashes of low mass AGB stars. Examples dealing 
with much longer time scales are discussed by R. Reifarth 
and A. Mengoni in their contributions to this volume. 

To decipher the information contained in abundance patterns
can be an intricate and complex process, which requires \\
- accurate nuclear physics data, in particular the rates 
for production and transmutation by nuclear reactions, but 
often also nuclear structure properties for describing the 
impact of temperature, and \\
- sufficiently detailed astrophysical prescriptions for a 
realistic or at least plausible modeling of the relevant
phenomena. It is this duality that we had the pleasure to 
pursue in the Torino-FZK liaison for more than two decades, 
and which, in return, was found to provide surprising 
constraints for the physics of stellar evolution.

\newcommand{\noopsort}[1]{} \newcommand{\printfirst}[2]{#1}
  \newcommand{\singleletter}[1]{#1} \newcommand{\swithchargs}[2]{#2#1}

\end{document}